\documentstyle[11pt]{article}
\begin{document}
\centerline {\bf Fission and cluster decay of $^{76}Sr$ nucleus in the
groud-state and formed in} 
\vskip 6pt
\centerline {\bf heavy-ion reactions}
\vskip 12pt
\centerline {R.K. Gupta$^{1,2)}$, M.K. Sharma$^{1,2}$, S. Singh$^{1)}$,
R. Nouicer$^{2)}$ and C. Beck$^{2)}$} 
\vskip 6pt
\centerline {1) Physics Department, Panjab University, Chandigarh-160014, India}
\centerline {2) Institut de Recherches Subatomiques, F-67037 Strasbourg Cedex
2, France} 
\vskip 24 pt
\par\noindent
{\bf Abstract:} Calculations for fission and cluster decay of $^{76}Sr$ are
presented for this nucleus to be in its ground-state or formed as an excited
compound system in heavy-ion reactions. The predicted mass distribution, for
the dynamical collective mass transfer process assumed for fission of
$^{76}Sr$, is clearly asymmetric, favouring $\alpha $-nuclei. Cluster decay is
studied within a preformed cluster model, both for ground-state to ground-state
decays and from excited compound system to the ground-state(s) or excited
states(s) of the fragments. 
\vskip 24pt
\baselineskip 24pt
$^{76}_{38}Sr$ is a superdeformed nucleus with an estimated quadrupole
deformation $\beta _2=0.44$ (see Fig. 1 in [1]). From the point of view of
known {\bf spherical} shell closures at Z=N=40, such a large ground-state
deformation for $^{76}Sr$ means the natural breaking of these spherical shells
and hence nuclear instability against both fission and exotic cluster decay
processes. However, we shall see in the following that, like the other
superdeformed nuclei in this mass region [1], though this nucleus is naturally
stable (negative Q-value) against only light clusters with masses $A_2<12$, the
calculated cluster decay half lives for $A_2\ge 12$ are also large enough
($T_{1/2}>10^{80} s$) to term this nucleus  as a stable nucleus against all
cluster decays. This kind of stability could apparently be due to stable {\bf
deformed} shell closures at Z=N=38, predicted earlier in many other
calculations [2,3]. Alternatively, if these nuclei are prepared in heavy-ion
collisions, then, depending on the excitation energy of the compound nucleus
formed, both fission  (also called, fusion-fission) and cluster decay are the
viable processes. The present day experiments are directed at these studies
(see e.g. [4-7] and earlier references there in). 
\par
It now seems accepted that compound systems with $A\le 42$ are characterized by
nuclear orbiting phenomena (the deep inelastic process), although a considerable
amount of yield due to fusion-fission could not be ruled here too [8,9]. On the
other hand, the systems with $A\approx 47-60$ are  strongly the cases of
fusion-fission process (fully energy-damped fragments) since for all the cases
studied so-far the observed yields are independent of nuclei in the entrance
channel and no strong peaking of yields is observed near the target and
projectile masses [4-7]. In all these cases, asymmetric mass splitting is
favoured and hence lie far below the Businaro-Gallone transition point [10]
(the fissility parameter $x(=Z^2/50A$) is less than the $x_{BG}=0.396$ for
$\ell =0$ and this value decreases as $\ell$-value increases). For nuclei with
$A\sim 80$, the situation is not so clear. Only three experiments are made
[11,12] that form the compound systems $^{78}_{38}Sr$, $^{80}_{40}Zr$ and
$^{83}_{36}Kr$ and one calculation is available for $^{80}Zr$ [13]. Notice that
fissility $x=$0.370 and 0.312 ($<x_{BG}$), respectively, for $^{78}Sr$ and
$^{83}Kr$ but $x=$0.40($>x_{BG}$) for $^{80}Zr$. Also, at least $^{78}Sr$ and
$^{80}Zr$ are superdeformed in their respective ground-states. The interesting
result is: whereas the measured mass spectra of $^{83}Kr$ [11] and $^{80}Zr$
[12] are clearly asymmetric and symmetric, as expected, respectively, but that
of $^{78}Sr$ [12] is more asymmetric than symmetric.  The last experiment on
$^{28}Si+^{50}Cr$ forming the compound system $^{78}Sr^{\ast}$ is made only at
one energy ($E_{lab}=150 MeV$) and at one angle ($\theta _{lab}=30^0$). In view
of this result we have chosen $^{76}Sr$ nucleus with $x$(=0.38) as well as
deformation $\beta $ larger than that for $^{78}Sr$ (estimated $\beta _2=0.41$
for $^{78}Sr$). Such a study has not yet been taken up either experimentally or
theoretically. Our calculations in the following show a clear asymmetric mass
distribution, as expected from the point of view of Businaro-Gallone transition
[10]. This result suggests that a properly angle intergrated mass distribution
for $^{78}Sr$ (not yet measured) should also be clearly asymmetric since $x$
for $^{78}Sr$ is smaller than that for $^{76}Sr$. 
\par
Theoretically, both the quantum mechanical fragmentation process [13-16] as
well as the statistical models [8,11,17-19] have been used to explain the
measured mass distributions in these reactions. The statistical or compound
nucleus model calculations, assuming fusion-fission process, are made for three
possible cases of (i) two spheres separated by a fixed distance d=2fm [11],
(ii) saddle point shapes, called transition state model [8], and (iii) scission
point shapes with d taken as a variable [17-19]. In this last case, the
Hauser-Feshbach formalism  [20] is extended to include the decay fragments
heavier than $\alpha$-particle. Perhaps, the preformation factor for different
fragments should also be added to this extended Hauser-Feshback method (EHFM).
In a statistical model all open decay channels are taken to be equally
populated. This is true as long as one is talking of $\gamma $-decay and the
light particle evaporation of n, p and $\alpha $-particle. However, once the
heavy fragments are also included, $\alpha $-particle is known to compete with
some heavy fragments (the exotic cluster decays) and hence, a preformation
probability factor between $\alpha $- and the other heavy-fragments should come
in. 
\par
In this paper, we use the quantum mechanical fragmentation theory, the QMFT
[13-16], and the cluster decay model [1,21,22] based on QMFT. According to the
QMFT, the binary fragmentation is a collective mass transfer process where both
the light and heavy  fragments (including the light particles) are produced
with different quantum mechanical probabilities. Applications of both the
fragmenation and cluster decay processes are made in only a few cases for the
light systems [13-16] and infact it was on the basis of this theory that
fusion-fission was first proposed by Gupta and collaborators in 1984 as the
possible explanation of the observed data on fragmentation of light systems. 
\par
In the QMFT, a dynamical collective coordinate of mass (and charge) asymmetry
$\eta={{A_1-A_2}\over {A_1+A_2}}$ (and $\eta _Z={{Z_1-Z_2}\over {Z_1+Z_2}}$) is
introduced whose limiting values are 0 and 1 i.e. $0\le \eta \le 1$ [14,23-27].
Since the potentials $V(R,\eta )$ and $V(R,\eta _Z)$, calculated within the
Strutinsky renormalization procedure ($V=V_{LDM}+\delta U$) by using the
appropriate liquid drop model (for $V_{LDM}$) and the asymmetric two-centre
shell model (for $\delta U$), are nearly independent of the relative separation
coordinate R, R can be taken as a time-independent parameter and hence solve
the stationary (instead of time-dependent) Schr\"odinger equation in $\eta$: 
$$\{ -{{\hbar^2}\over {2\sqrt B_{\eta \eta}}}{\partial \over {\partial
 \eta}}{1\over {\sqrt B_{\eta \eta}}}{\partial\over {\partial \eta }}+V_R(\eta
 )\} \psi _R^{(\nu )}(\eta ) = E_R^{(\nu )} \psi _R^{(\nu )}(\eta ).\eqno (1)$$
The R-value for light nuclei is fixed at the touching configuration of two
nuclei [13-16]:
$$R=R_1+R_2\eqno (2a)$$
with
$$R_i=1.28A_i^{1/3}-0.76+0.8A_i^{-1/3}.\eqno (2b)$$
In this approximation, the fragmentation potential 
$$V(\eta )=-B_1(A_1,Z_1) -B_2(A_2,Z_2)+E_c+V_P, \eqno (3)$$ 
where $B_i(A_i,Z_i)$ are the experimental binding energies [28], $E_c$ =
Z${_1}$Z${_2}$e${^2}$/R and $V_P$ is the additional attraction due to nuclear
proximity potential, given by the well known pocket formula of Blocki et. al
[29]. The charges $Z_i$ in (3) are fixed by minimizing $V(\eta _Z)$, defined by
(3) without $V_P$, in $\eta _Z$-coordinate. The rotational energy due to
angular momentum ($V_l$) is not added here since its contribution to the
structure of yields is shown to be small for lighter systems [15]. Of course,
$V_l$ should be added for a comparison of the relative yields. It may be
mentioned here that in a more realistic calculation, the two-centre nuclear
shape should be used, instead of eq. (2). One can then trace the actual nuclear
shapes involved. For a more quantitative comparison, perhaps $R=R_1+R_2+d(\le
2$fm) would be a better choice because then one is closer to the saddle shape. 
\par
The numerical solution of (1), on proper scaling, gives the fractional mass
distribution yields for each fragment as
$$Y(A_i)=\mid \psi _R(\eta (A_i))\mid ^2 {\sqrt {B_{\eta \eta}}} {2\over A},
 \eqno (4)$$
(i= 1 or 2). For the mass parameters $B_{\eta \eta}$ we use the classical
hydrodynamical model of Kr\"oger and Scheid [30]. For two touching spheres,
this model gives a simple analytical expression,
$$B_{\eta \eta}={{AmR_t^2}\over 4}{({{v_t(1+\beta )}\over {v_c}}-1)}\eqno (5a)$$
with
$$\beta ={R_c\over {4R_t}}{(2-{R_c\over R_1}-{R_c\over R_2})}\eqno (5b)$$
$$v_c=\pi R_c^2 R_t,\qquad R_c=0.4R_2\eqno (5c)$$
and $v_t=v_1+v_2$, the total conserved volume. Also, $R_2<<R_1$ and $R_c(\neq
0)$ is the radius of a cylinder of length $R_t$, whose existence allows a
homogenous, radial flow of mass between the two fragments. Here m is the
nucleon mass. 
\par
The nuclear temperature effects in (4) are also included through a
Boltzmann-like function 
$$\mid \psi _R\mid ^2=\sum _{\nu =0}^{\infty}\mid \psi _R^{(\nu )}\mid ^2 exp
 (-E_R^{(\nu )}/\theta )\eqno (6)$$
with $\theta$, the nuclear temperature in MeV, related to the excitation energy
as
$$E^{\ast }={1\over 9}A{\theta }^2-\theta .\eqno (7)$$
Furthermore, in some of the calculations here, temperature effects are taken to
act also on the shell effects as follows [31]
$$V=V_{LDM}+\delta U exp (-{\theta }^2/2.25). \eqno (8)$$
Similarly, mass parameters should also vary with temperature but no useable
prescription is available todate. A constant average mass is taken to mean a
complete washing of shell effects in it. 
\par
For cluster-decay calculations, we use the preformed cluster model (PCM) of
Malik and Gupta [21,22]. This model, based on the QMFT, also uses the decoupled
approximation to R- and $\eta $-motions and define the decay half-life
$T_{1\over 2}$ or the decay constant $\lambda $ as 
$$\lambda ={{{ln 2}\over {T_{1\over 2}}}}=P_0\nu P. \eqno (9)$$
Here, $P_0$ is the cluster preformation probability at a fixed R, given by the
solution of the stationary Schr\"odinger equation (1). At $R=R_1+R_2$,
$P_0=Y(A_2)$, given by eqs. (4) and (6). For ground-state to ground-state
decay, $\nu =0$. In the following, we choose $R=R_1+R_2$ (instead of $R=R_0$,
the compound nucleus radius, where $V(R_0)=Q$-value) since this assimilates the
effects of both deformations of the two fragments and neck formation between
them [32]. 
\par
P is the tunnelling probability, which can be obtained by solving the
corresponding stationary Schr\"o- dinger equation in R. Instead, Malik and
Gupta calculated it as the WKB penetrability which for the tunneling path shown
in figure 1 of [22] is given by 
$$P=P_i P_b \eqno (10a)$$
with
$$P_i=exp(-{2\over \hbar}{{\int }_{R_1+R_2}^{R_i}\{ 2\mu [V(R)-V(R_i)]\} ^{1/2}
 dR} \eqno (10b)$$
$$P_b=exp(-{2\over \hbar}{{\int }_{R_i}^{R_b}\{ 2\mu [V(R)-Q]\} ^{1/2} dR}.
 \eqno (10c)$$
This means that tunneling begins at $R=R_1+R_2$ and terminates at $R=R_b$ with
$V(R_b)=Q$. The de-excitation probability between $P_i$ and $P_b$ is taken to
be unity here. Both (10b) and (10c) are solved analytically [21,22].
Apparently, we are considering here the decay from the ground-state of the
parent nucleus to the ground-states of the decay products. On the other hand,
if the compound system is excited or the system ends in the excited state of
one or both the decay products, the Q-value has to be adjusted accordingly (as
discussed in the following). 
\par
In eq. (9), $\nu $ is the assault frequency, given simply as
$$\nu = {{velocity}\over {R_0}}=(2E_2/\mu )^{1/2}/R_0, \eqno (11)$$
where $E_2=(A_1/A) Q$ is the kinetic energy, taken as the Q-value shared
between two fragments, and $\mu =m({{A_1A_2\over A}})$ is the reduced mass. 
\par
Figure 1 shows the fragmentation potential for $^{76}Sr$, plotted as a function
of the light fragment mass $A_2$. Calculations are made in steps of one nucleon
transfer. We notice in Fig. 1 that potential energy minima lie only at N=Z,
A=4n nuclei, showing the strong shell effects of $\alpha $-nuclei. It is
important to realize here that this fragmentation potential is independent of
the nuclei in the entrance channel and the light ($A_2$) and heavy
($A_1=A-A_2$) fragments occur in coincidence. 
\par
The calculated fractional yields are shown in Fig. 2 for the fission of
$^{76}Sr$ from ground-state ($\nu =0$) and at an arbitrary $\sim 32$ MeV of
excitation energy ($\theta =2$ MeV). We notice that in the ground-state, almost
all the yield is taken away by the $\alpha $-particle alone, but as the
compound system is heated up, other fragments also show up. The interesting
results are (i) the yields are largest for $\alpha $-nuclei, and (ii) the mass
distribution is strongly asymmetric. The fact that both the above mentioned
results also hold good in the very small yield zone of symmetric fragmentation,
we have shown in Fig. 3, the renormalized fractional yields, calculated for the
fragment masses $15\le A_i\le 61$ only. Apparently, the $\alpha $-nuclei are
favoured and the overall mass distribution is asymmetric. Also, the role of
temperature in increasing the yields, more so for the symmetric and nearly
symmetric fragments, is also shown in Fig. 3. 
\par
We have also analyzed the role of temperature on shell effects $\delta U$. For
this purpose, we have redone our calculations by using the theoretical binding
energies [33] where $V_{LDM}$ and $\delta U$ contributions are tabulated
separately. The calculated fragmentation potentials at different temperatures
and the resulting yields are shown, respectively, in Figs. 4 and 5. We notice
that temperature effects are large but the general character of the mass
distribution remains unaffected. Our earlier calculations [15,16] show that the
contribution of shell effects in mass parameters $B_{\eta \eta }$ are also of
similar orders and act in the same way i.e. without disturbing the general
character of the mass distribution. Following [16], the small entrance channel
effects in the observed mass distributions could be assimilated by the
empirically fitted $B_{\eta \eta }(\eta )$. 
\par
Our calculated decay half-lives for the decay of $^{76}Sr$ from its
ground-state to the ground-states of all $\alpha $-nuclei clusters with
positive Q-value are given in Table 1. Of course, the decay could also occur
into the excited states of the daughter and/or cluster nuclei. This would mean
decreasing Q-value to $Q-E_i$, where, say, $E_i$ is the energy of first excited
state of daughter nucleus. This type of decay is studied in the following
paragraph (Table 2). We notice in Table 1 that the penetrabilities P are very
small. This is particularly so for $^{12}C$ decay since its Q-value is
relatively very small. Apparently, all the predicted decay half-lives are very
large ($T_{1/2}>10^{80} s$) and $^{76}Sr$ nucleus can be said to be stable
against all possible cluster decays. Another interesting point to note is that
the daughter nuclei for $^{20}Ne$- and $^{32}S$-decays are the doubly magic N=Z
nuclei. Hence, in view of the so-far observed cluster decays [34] , these two
decays should be the most probable decays. However, the preformation factors
$P_0$ for both $^{20}Ne$ and $^{32}S$ clusters are also very small, though the
decrease of $P_0$ with the increase of cluster size is very much in agreement
with the situation for observed decays (see e.g. Fig. 13 in the review [34]). 
\par
The role of excitation energy $E^{\ast }$ of the compound system is presented
in Table 2. Once again the decays could end into the ground-states or excited
states $E_i$ of the fragments. For decay into ground-state(s) of the
fragment(s), the effective Q-value ($Q_{eff}$) will become $Q+E^{\ast }$ but if
it goes into an excited state $E_1$ of one fragment, $Q_{eff}=Q+E^{\ast }-E_1$.
Both the cases are studied in Table 2 for some arbitrary values of $E^{\ast }$
and the first excited states of the daughter fragments. Interesting enough, the
decay constant  $\lambda $ increases (or $T_{1/2}$ decreases) considerably.
Such calculations for light nuclei are made for the first time and have become
relevant because it is now possible to study experimentally [35] the above
mentioned fine structure effects of decay into the excited states of fission
fragments. Once the data becomes available, it will also help deciding between
the fission and cluster decays of these light nuclear systems. 
\par
Summarizing, we have presented our calculations for fission and cluster decays
of $^{76}Sr$. Both the cases of $^{76}Sr$ in the ground-state and produced as
an excited compound system in heavy-ion reactions are studied. Here fission is
treated as a collective mass transfer process and cluster decay studies are
based on a model allowing preformation of clusters. Calculations show that, in
a clear asymmetric mass distribution, $^{76}Sr$ nucleus allows preferential
$\alpha $-nuclei transfer resonances as well as decays. The only experimental
study available on this system is $1\alpha $ and $2\alpha $ transfer products
from $^{36}Ar$ on $^{40}Ca$ at an energy near the Coulomb barrier ($V_c=53.6$
MeV) [36]. Our calculations suggest that for obtaining the complete fission
products, one has perhaps to go to at least double the Coulomb barrier
energies. Also, following Sobotka et. al [11], use of inverse kinematics
(projectile heavier than the target) may be an additional help. This provides a
large center-of-mass velocity which facilitates the verification of full
momentum transfer and easy identification of the fragment's atomic number at
higher incident energies. Also, the high-enegy solution at forward angles
should enhance the observation of compound-nucleus decay and virtually
eliminate any possible deep-inelastic contribution. 
\vskip 24pt
One of us (RKG) would like to express his thanks to Professor Dr. Bernard Hass,
Director CRN, Strasbourg, and his colleagues for the nice hospitality of the
Institute extended to him for a month during the autumn of 1996 when this work
was completed. Brief reports of this work were contributed earlier at the
Second Int. Conf. on Atomic and Nuclear Clusters 1993, Santorini (Greece), June
28- July 2, 1993, and at Cluster 94: Clusters in Nuclear Structure and
Dynamics, Strasbourg (France), Sept. 6-9, 1994. 
\vfill\eject
\par\noindent
{\bf References}
\vskip 6pt
\begin{enumerate}
\begin{description}
\item {1.} R.K. Gupta, W. Scheid and W. Greiner, J. Phys. G: Nucl. Part. Phys.
 {\bf 17} (1991) 1731
\item {2.} P. M\"oller and J.R. Nix, Nucl. Phys. A {\bf 361} (1981) 117; At.
 Data Nucl. Data Tables {\bf 26} (1981) 165.
\item {3.} R. Bengtsson, P. M\"oller, J.R. Nix and J.-Y. Zhang, Phys. Scr. {\bf
 29} (1984) 402
\item {4.} C. Beck, D. Mahboub, R. Nouicer, T. Matsuse, B. Djerroud, R.M.
 Freeman, F. Hass, A. Hachem, A. Morsad, M. Youlal, S.J. Sanders, R. Dayras,
 J.P. Wieleczko, E. Berthoumieux, R. Legrain, E. Pollacco, Sl. Cavallaro, E. De
 Filippo, G. Lanzano, A. Pagano and M.L. Sperduto, Phys. Rev. C {\bf 54} (1996)
 227
\item {5.} K.A. Farrar, S.J. Sanders, A.K. Dummer, A.T. Hasan, F.W. Prosser,
 B.B. Back, I.G. Bearden, R.R. Betts, M.P. Carpenter, B. Crowell,  M. Freer,
 D.J. Henderson, R.V.F. Janssens, C. Beck, R.M. Freeman, Sl. Cavallaro and A.
 Szanto de Toledo, Phys. Rev. C {\bf 54} (1996) 1249
\item {6.} S.J. Sanders, A.Hasan, F.W. Proser, B.B. Back, R.R. Betts, M.P.
 Carpenter, D.J. Henderson, R.V.F. Janssens, T.L. Khoo, E.F. Moore, P.R. Wilt,
 F.L.H. Wolfs, A.H. Wuosmaa, K.B. Beard and Ph. Benet,  Phys. Rev. C {\bf 49}
 (1994) 1016
\item {7.} Sl. Cavallaro, C. Beck, E. Berthoumieux, R. Dayras, E. De Filippo, G.
 Di Natale, B. Djerroud, R.M. Freeman, A. Hachem, F. Hass, B. Heusch, G.
 Lanzano, R. Legrain, D. Mahboub, A. Morsad, A. Pagano, E. Pollacco, S.J.
 Sanders and M.L. Sperduto, Nucl. Phys. A {\bf 583} (1995) 161
\item {8.} S.J. Sanders, Phys. Rev. C {\bf 44} (1991) 2676
\item {9.} N. Aissaoui, F. Hass, R.M. Freeman, C. Beck, M. Morsad, B. Djerroud,
 R. Caplar and A. Hachem,  Phys. Rev. C (1996) to be published
\item {10.} U.L. Businaro and S. Gallone, Nuo. Cim. {\bf 1} (1955) 629, 1277
\item {11.} L.G. Sobotka, M.A. McMahan, R.J. McDonald, C. Signarbieux, G.J.
 Wozniak, M.L. Padgett, J.H. Gu, Z.H. Liu, Z.Q. Yao and L.G.Moretto, Phys. Rev.
 Lett. {\bf 53} (1984) 2004
\item {12.} P.M. Evans, A.E. Smith, C.N. Pass, L. Stuttge, B.B. Back, R.R.
 Betts, B.K. Dichter, D.J. Henderson, S.J. Sanders, F. Videback and B.D.
 Wilkins, Phys. Lett. B {\bf 229} (1989) 25; Nucl. Phys. A {\bf 526} (1991) 365
\item {13.} R.K. Puri and R.K. Gupta, J. Phys. G: Nucl. Phys. {\bf 18} (1992)
 903;
\item { } R.K. Puri, S.S. Malik and R.K. Gupta, Europhys. Lett. {\bf 9} (1989)
 767
\item {14.} R.K. Gupta, D.R. Saroha and N. Malhotra, J. de Physique Coll. {\bf
 45} (1984) C6-477
\item {15.} D.R. Saroha, N. Malhotra and R.K. Gupta, J. Phys. G: Nucl. Phys.
 {\bf 11} (1985) L27
\item {16.} S.S. Malik and R.K. Gupta, J. Phys. G: Nucl. Phys. {\bf 12} (1986)
 L161
\item {17.} T. Matsuse, S.M. Lee, Y.H. Pu, K.Y. Nakagawa, C. Beck and T.
 Nakagawa, in {\it Towards a Unified Picture of Nuclear Dynamics}, Nikko, June
 1991, ed. Y. Abe, S.M. Lee and F. Sakata, AIP Conf. Proc. No. 250, p.112
\item {18.}  R. Nouicer, C. Beck, D. Mahboub, T. Matsuse, B. Djerroud, R.M.
 Freeman, A. Hachem, Sl. Cavallaro, E. De Filippo, G. Lanzano, A. Pagano, R.
 Dayras, E. Berthoumieux, R. Legrain and E. Pollacco, Z. Phys. {\bf A356} 
(1996) 5
\item {19.} T. Matsuse, C. Beck, R. Nouicer and D. Mahboub, Phys. Rev. C
{\bf 55} (1997) 1380
\item {20.} W. Hauser and H. Feshbach, Phys. Rev. {\bf 87} (1952) 366
\item {21.} R.K. Gupta, Proc. 5th Int. Conf. on Nuclear Reaction Mechanisms,
 Varenna, 1988, p.416
\item {22.} S.S. Malik and R.K. Gupta, Phys. Rev. C {\bf 39} (1989) 1992
\item {23.} H.J. Fink, W. Greiner, R.K. Gupta, S. Liran, H.J. Maruhn, W. Scheid
 and O. Zohni, Proc. Int. conf. on Reactions between Complex Nuclei, Nashville,
 Tenn. USA, June 10-14, 1974, eds. R.L. Robinson, F.K. McGowan, J.B. Ball and
 J.M. Hamilton, North Holland Pub. 1974, Vol. 2, p.21
\item {24.} J. Maruhn and W. Greiner, Phys. Rev. Lett. {\bf 32} (1974) 548
\item {25.} R.K. Gupta, W. Scheid and W. Greiner, Phys. Rev. Lett. {\bf 35}
 (1975) 353
\item {26.} R.K. Gupta, Sovt. J. Part. Nucleus {\bf 8} (1977) 289
\item {27.} J.A. Maruhn, W. Greiner and W. Scheid, {\it Heavy Ion Collisions},
 Vol. 2, ed. R. Bock, North Holland: Amsterdam, Ch. 6
\item {28.} G. Audi and A.H. Wapstra, Nucl. Phys. A {\bf 595} (1995) 409
\item {29.} J. Blocki, J. Randrup, W.J. Swiatecki and C.F. Tsang, Ann. Phys.
 (NY) {\bf 105} (1977) 427
\item {29.} H. K\"oger and W. Scheid, J. Phys. G: Nucl. Phys. {\bf 6} (1980) L85
\item {30.} A.S. Jensen and J. Damgaard, Nucl. Phys. A {\bf 203} (1973) 578
\item {31.} S. Kumar and R.K. Gupta, Phys. Rev. C (1996)- in press;
\item { } R.K. Gupta, Proc. NATO Advanced Study Institute, {\it Frontier Topics
 in Nuclear Physics}, Predeal, Romania, Aug. 24- Sept. 4, 1993, eds. W. Scheid
 and A. S\v andulescu, Plenum Pub. Corp. 1994, p.129
\item {32.} P.E. Haustein, At. Data Nucl. Data Tables {\bf 39} (1988) 185
\item {33.} See e.g. the review:
\item { } R.K. Gupta and W. Greiner, Int. J. Mod. Phys. E {\bf 3} (1994, Suppl.)
 353
\item {34.} R. Nouicer, Private communication
\item {35.} T. Kirchner, Ph.D. Thesis 1994, Freien Universit\"at, Berlin,
 Germany.

\end{description}
\end{enumerate}
\vfill\eject
\par\noindent
{\bf Figure Captions:}
\vskip 6pt
\begin{enumerate}
\begin{description}
\item {Fig. 1.}  The mass fragmentation potential for $^{76}Sr$, calculated by
 using the experimental binding energies. Only the light fragments $A_2$ are
 shown along the x-axis. The other fragment $A_1=A-A_2$.
\item {Fig. 2.} The calculated mass distribution yields for the fission of
 $^{76}Sr$ from the ground-state and at a temperature $\theta =2$ MeV ($E^{\ast
 }\approx 32$ MeV). The range of fragment masses is $1\le A_i\le 75$.
\item {Fig. 3.} The same as for Fig. 2 but at $\theta =2$ and 3 MeV and for the
 range $15\le A_i\le 61$. The fragmentation potential used is the same as in
 Fig. 1 but for the range $15\le A_i\le 61$ only.
\item {Fig. 4.} The same as for Fig. 1 but by using the theoretical binding
 energies and for the mass range $15\le A_i\le 61$ only. Also, the temperature
 effects on shell corrections $\delta U$ are shown for $\theta =2$ and 3 MeV.
\item {Fig. 5.} The same as for Fig. 3 but by using the potentials of fig. 4.

\end{description}
\end{enumerate}

\end{document}